\documentclass[preprint,authoryear,12pt]{elsarticle}

\usepackage{graphics}
\usepackage{amssymb}
\usepackage[ps2pdf,%
a4paper=true,%
breaklinks=true,%
colorlinks=true,%
pdfauthor={First Author et al.},%
pdftitle={Template for manuscripts in Advances in Space Research}%
]{hyperref}
\journal{Advances in Space Research}
\renewcommand{\vec}[1]{\mbox{\boldmath $#1$}}
\def\lsim{\lower.4ex\hbox{$\;\buildrel <\over{\scriptstyle\sim}\;$}}
\def\gsim{\lower.4ex\hbox{$\;\buildrel >\over{\scriptstyle\sim}\;$}}

\def\aap{A\&A}
\def\apj{ApJ}
\def\apss{Ap\&SS}
\def\mnras{MNRAS}
\def\solphys{Sol. Phys.}
\begin{document}

\begin{frontmatter}

\title{Diamagnetic pumping in a rotating convection zone}

\author[label1,label2]{L.\,L.~Kitchatinov\corref{cor}}
\cortext[cor]{Corresponding author}
\ead{kit@iszf.irk.ru}
\author[label1]{A.\,A.~Nepomnyashchikh}
\ead{nep\underline{ }a@iszf.irk.ru}
\address[label1]{Institute for Solar-Terrestrial Physics, Lermontov Str. 126A, Irkutsk 664033, Russia}
\address[label2]{Pulkovo Astronomical Observatory, St. Petersburg 196140, Russia}
\begin{abstract}
Solar dynamo models require some mechanism for magnetic field concentration near the base of the convection zone in order to generate super-kilogauss toroidal fields with sufficiently large ($\sim 10^{24}$\,Mx) magnetic flux. We consider the downward diamagnetic pumping near the base of the convection zone as a possible concentration mechanism and derive the pumping velocities with allowance for the effect of rotation. Transport velocities for poloidal and toroidal fields differ in rotating fluid. The toroidal field is transported downward along the radius only but the pumping velocity for the poloidal field has an equatorward meridional component also. Previous results for cases of slow and rapid rotation are reproduced and the diamagnetic pumping expressions adapted for use in dynamo models are presented.
\end{abstract}

\end{frontmatter}

\parindent=0.5 cm

\section{Introduction}
Large-scale magnetic fields of turbulent fluids do not strictly follow the mean fluid motion. This is not only because the fields are subject to turbulent diffusion. Mean-field magnetohydrodynamics also predicts the effects of the large-scale field pumping with effective velocities which are not the actual velocities of fluid motion. The fields are expelled from the regions of relatively high turbulence intensity with the effective velocity $\vec{V}_\mathrm{dia} = -{\vec\nabla}\eta_{_\mathrm{T}}/2$ \citep[$\eta_{_\mathrm{T}}$ is the eddy magnetic diffusivity;][]{KR80}. Diamagnetic pumping was also found in 3D numerical experiments \citep{Tea98,Tea01,DN01,Oea02,ZR03,Kea06} and in a laboratory experiment with liquid sodium \citep{Sea07}.

Diamagnetic pumping is significant for the modelling of the solar dynamo. \citet{KKT06} and \citet{GG08} found that allowance for the pumping brings their models' results closer to observations. They in particular noticed that a horizontal component, which the pumping velocity may possess in addition to its (dominant) radial part, can be significant for latitudinal drift of magnetic fields. The observed equatorial drift of sunspot activity is believed to result primarily from a deep equatorward meridional flow \citep{CSD95,D95}. However, horizontal turbulent pumping - if it exists - may also play a role.

Horizontal pumping can result from the influence of rotation on convective turbulence. The influence elongates convective eddies along the rotation axis to induce turbulence anisotropy. The resulting {\em anisotropic} pumping not only achieves a horizontal component but transports the poloidal and toroidal large-scale fields with different velocities. These circumstances were demonstrated in the case of effective transport due to stratification of density in turbulent fluid \citep{K91} - a weaker effect compared to the diamagnetic pumping near the base of convection zone. Almost all effects of mean-field MHD for rotating turbulence were analysed in the nineties except diamagnetic pumping \citep[cf., however,][]{P08}. This paper fills the gap.

We shall see that the diamagnetic pumping indeed attains a horizontal component in rotating fluids but only for poloidal fields which are transported with a different velocity compared to that of toroidal fields. The relative value of toroidal field transport (in relation to the turbulent diffusion) is enhanced by rotation. Pictorial interpretations can be given to all of these findings. Before deriving the diamagnetic pumping, however, it is worth discussing in more detail why can it be significant for the solar dynamo.
\section{Significance of diamagnetic pumping}
The region near the base of the convection zone has long been  recognised as a favourable site for the solar dynamo. Diamagnetic pumping concentrates magnetic fields towards this site. The generation of super-kilogauss toroidal fields is hardly possible without this concentration.

\begin{figure}[th]
\centering
   \includegraphics[width=10 truecm, angle=0]{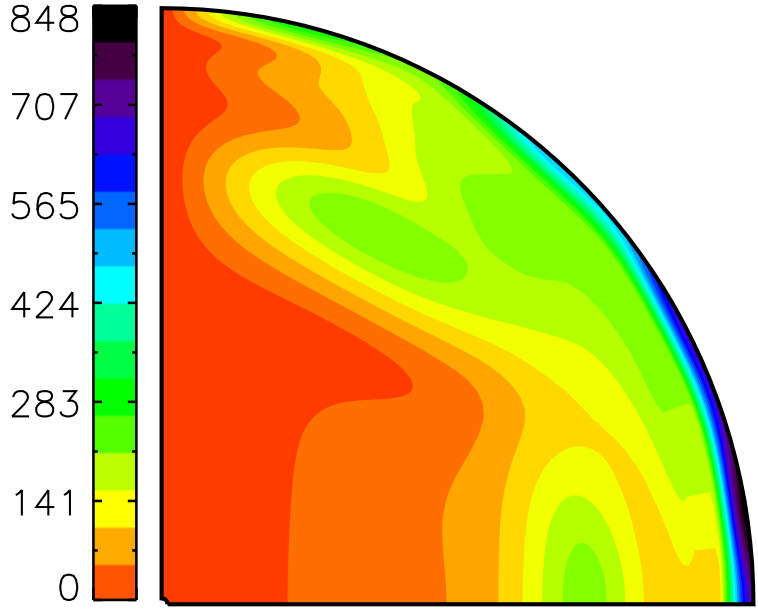}
   \caption{The poloidal-to-toroidal field conversion factor of Eq.\,(\ref{2})
            estimated from the helioseismological data on the gradients of rotation rate of \citet{ABC08} (colour online).
            }
   \label{f1}
\end{figure}

It is generally acknowledged that toroidal fields ($B_\phi$) are wound from poloidal ones (${\vec B}^\mathrm{p}$) by differential rotation. The winding process is accounted for by the following term,
\begin{equation}
    \frac{\partial B_\phi}{\partial t} = r\sin\theta ({\vec B}^\mathrm{p}\cdot{\vec\nabla})\Omega  + . . . ,
    \label{1}
\end{equation}
on the right-hand side of the toroidal field induction equation. In this equation, $\Omega$ is the angular velocity, $r$ is the heliocentric radius and $\theta$ is the co-latitude. The upper bound for the amplitude of the toroidal field, which can be wound in the course of a solar cycle, can be estimated as $B_\phi = C B^\mathrm{p}$, where $B^\mathrm{p}$ is the (local) amplitude of the poloidal field and the conversion factor $C$ reads
\begin{equation}
    C = r\sin\theta|{\vec\nabla}\Omega| P_\mathrm{cyc} ,
    \label{2}
\end{equation}
with $P_\mathrm{cyc} = 11$\,years as a cycle period. The angular velocity gradients in the solar interior were detected seismologically by \citet{ABC08}. The conversion factor of Eq.\,(\ref{2}) estimated using their data (GONG data averaged over the 23rd solar cycle) is shown in Fig.\,\ref{f1}.

Helioseismology may underestimate the radial gradient of rotation in the tachocline region. However, the radial gradient shears the radial field, which should be small in the tachocline region. Otherwise it is a relic field penetrating from the radiative interior. The conversion factor of Eq.\,(\ref{2}) is an upper bound, which can be reached only if the poloidal field and rotation gradient are perfectly aligned. Also taking $P_\mathrm{cyc}$ as the winding time overestimates the conversion factor. The toroidal field is amplified only on the rise phase of an activity cycle, before the poloidal field inversion. However, even the upper bounds of the conversion factor of Fig.\,\ref{f1} are too small for producing super-kilogauss toroidal fields with distributed dynamo models.

\citet{GW81} estimated the total magnetic flux of the active regions of a solar cycle to exceed $10^{24}$\,Mx. Their estimations also suggest that a comparable or perhaps slightly smaller toroidal flux should be stored near the bottom of the convection zone at the activity maxima. Assuming the latitudinal extend of the storage region of about $30^\circ$ and its radial extend $< 0.1R_\odot$, we find the low bound for the  toroidal field strength of about $4000$\,Gs. \citet{CS15} found that the net toroidal flux does not depend on how the poloidal field is distributed inside the convection zone. The flux is uniquely defined by the surface radial field and differential rotation. The observed surface fields correspond to the peak values of toroidal flux of about $5\times 10^{23}$\,Mx in the recent solar cycles 22 and 23 \citep{CS15}. If the surface active regions emerge from near the base of the convection zone, super-kiligauss toroidal fields should be present in the near-base region.

The large-scale poloidal field on the solar surface is of the order of 1 Gauss \citep{S88,Oea06}. If the field is distributed smoothly inside the convection zone, toroidal fields well below one kilogauss are produced near its base (Fig.\,\ref{f1}). It may be argued that this refers to the mean or large-scale field only. The field in convectively-unstable fluids fragments in a fibril state with a much stronger field in the fibrils compared to its mean value \citep{P84}. However, the magnetic flux of a sub-kilogauss {\em mean} toroidal field in the near-bottom region is too small ($< 10^{24}$\,Mx) to feed active regions of a solar cycle.

The problem can be resolved by downward diamagnetic pumping. The meridional field pumped to the base of convection zone can be two orders of magnitude stronger compared to the surface poloidal field \citep{KO12}. The conversion factors of Fig.\,\ref{f1} then suffice for producing super-kilogauss toroidal fields. It may also be noted that the model with diamagnetic pumping realises an interface dynamo in spite of the model's being of distributed type. The model is, however, sensitive to details of the turbulent diffusivity profile, specification of which remains to be a source of uncertainties.

Diamagnetic pumping seems to be necessary for reproducing the observationally suggested combination of weak poloidal and strong toroidal fields in dynamo models. Otherwise, polar fields over one hundred Gauss are required for producing super-kilogauss toroidal fields \citep{DC99,MT15}.
\section{Derivation of diamagnetic pumping}
Mean-field magnetohydrodynamics includes the effect of turbulence in the mean magnetic field ($\vec B$) induction equation,
\begin{equation}
    \frac{\partial{\vec B}}{\partial t} = {\vec\nabla}\times
    \left( {\vec{\cal E}} + {\vec V}\times{\vec B}
    - \eta{\vec\nabla}\times{\vec B}\right) ,
    \label{3}
\end{equation}
via the mean electromotive force (EMF),
\begin{equation}
    {\vec{\cal E}} = \langle {\vec u}\times{\vec b}\rangle,
    \label{4}
\end{equation}
which is the mean of the cross-product of the fluctuating velocity, $\vec u$, and magnetic field, $\vec b$.
\subsection{Qusi-linear approximation}
We derive that part of the EMF that represents the diamagnetic pumping. The derivations are performed in the quasi-linear approximation known also as the second-order correlation approximation (SOCA) or the first-order smoothing (FOSA). The approximation is justified for the cases of small Reynolds numbers or short correlation times only. Solar convective turbulence does not belong to either of these cases. However, the existence of the limits, for which the quasi-linear approximation is valid, ensures this approximation from unphysical results. The approximation can be expected to remain valid by order of magnitude also beyond its justification limits. The expectation often - though not always - comes true \citep{Kea06,Kea14}. However, some disagreement about the validity of  quasi-linear approximation at the astrophysical case of high magnetic Reynolds number still remains.

Inhomogeneity of the mean field $\vec B$ is not essential for the contribution of the diamagnetic pumping, ${\vec{\cal E}}^\mathrm{dia}$, in the EMF. Derivations are therefore performed for the uniform mean field. Inhomogeneity of the turbulence intensity has, however, to be included. Derivations are performed in a rotating frame to include the effect of rotation. The influence of the mean magnetic field on turbulence is neglected \citep[the nonlinear diamagnetic pumping was considered by][]{KR92}.

The quasi-linear approximation for rotating inhomogeneous fluids was discussed in detail elsewhere \citep{RK93} and will be repeated here only briefly. The approximation neglects the nonlinear terms in the equations for fluctuating magnetic and velocity fields. The characteristic temporal and spatial scales of fluctuating fields are assumed small compared to the mean-field scales. Variations of the mean fields during the correlation time and over the correlation length of turbulence are neglected in the derivations of the diamagnetic pumping. It is then convenient to use the Fourier transform for the fluctuating velocity,
\begin{equation}
    {\vec u}({\vec r},t) = \int \hat{\vec u}({\vec k},\omega )
    \mathrm{e}^{\mathrm{i}({\vec r}\cdot{\vec k} - \omega t)}d{\vec k}d\omega ,
    \label{5}
\end{equation}
and the same for the fluctuating magnetic field. The transform simplifies matters by converting partial differential equations into algebraic equations to yield the following relations
\begin{eqnarray}
    \hat{u}_i(\vec{k},\omega ) &=&
    D_{ij}(\vec{k},\omega , \vec{\Omega})
    \hat{u}^0_j(\vec{k},\omega ),
    \nonumber \\
    \hat{\vec b}(\vec{k},\omega) &=&
    \frac{i(\vec{k}\cdot\vec{B})}{\eta k^2 - i\omega}\hat{\vec u}(\vec{k},\omega ) ,
    \label{6}
\end{eqnarray}
where repetition of subscripts signifies summation and $\hat{u}^0$ corresponds to the so-called \lq\lq original" turbulence not perturbed by rotation. The influence of rotation is included via the tensor
\begin{equation}
    D_{ij}(\vec{k},\omega , \vec{\Omega}) =
    \frac{\delta_{ij} + \hat{\Omega}\varepsilon_{ijl}k_l/k}{1 + \hat{\Omega}^2} ,
    \ \ \ \
    \hat{\Omega} = \frac{2(\vec{k}\cdot\vec{\Omega})}{k(\nu k^2 - i\omega)} .
    \label{7}
\end{equation}
The original turbulence is assumed to be statistically steady but not uniform. The simplest spectral tensor, which allows for the inhomogeneity of turbulence intensity, reads
\begin{eqnarray}
    \langle \hat{u}^0_i({\vec z},\omega)\hat{u}^0_j({\vec z}',\omega')\rangle &=&
    \frac{\hat{E}(k,\omega,{\vec\kappa})}{16\pi k^2} \delta(\omega+\omega')
    \nonumber \\
    &&\times\left( \delta_{ij} - k_ik_j/k^2
    + (\kappa_i k_j - \kappa_j k_i)/(2k^2)\right) ,
    \label{8}
\end{eqnarray}
where ${\vec k} = ({\vec z} - {\vec z}')/2$ and ${\vec\kappa} = {\vec z} + {\vec z}'$. $\hat{E}(k,\omega,{\vec\kappa})$ in Eq.\,(\ref{8}) is the Fourier transform of the local spectrum $E(k,\omega,{\vec r})$:
\begin{eqnarray}
    E(k,\omega,{\vec r}) &=& \int \hat{E}(k,\omega,{\vec\kappa})
    \mathrm{e}^{\mathrm{i}{\vec\kappa}\cdot{\vec r}}\mathrm{d}{\vec\kappa},
    \nonumber \\
    \langle {u^0}^2\rangle &=& \int\limits_0^\infty\int\limits_0^\infty
    E(k,\omega,{\vec r})\,\mathrm{d}k\,\mathrm{d}\omega .
    \label{9}
\end{eqnarray}

On using Eqs. (\ref{6}) to (\ref{8}), the Fourier transform of the EMF (\ref{4}) can be constructed. The inverse Fourier transform then gives the sought-for EMF. Only the even terms in the angular velocity should be kept in the derivations. The even terms represent the contribution ${\vec{\cal E}}^\mathrm{dia}$ of the diamagnetic pumping. The odd terms stand for the alpha-effect \citep{KR80}, which we do not consider.
\subsection{Anisotropic diamagnetic pumping}
The derivation of the diamagnetic pumping for rotating fluid outlined above yields
\begin{equation}
     {\vec{\cal E}}^\mathrm{dia} = {\vec U}^\mathrm{t}\times{\vec B} -
     \tilde{\vec U}\times{\vec e}\left({\vec e}\cdot{\vec B}\right) ,
     \label{10}
\end{equation}
where ${\vec e} = {\vec\Omega}/\Omega$ is the unit vector along the rotation axis and the effective velocities, ${\vec U}^\mathrm{t}$ and $\tilde{\vec U}$, are gradients of scalar functions:
\begin{equation}
    {\vec U}^\mathrm{t} = -{\vec\nabla}\eta_1,\ \ \
    \tilde{\vec U} = -{\vec\nabla}\eta_2.
    \label{11}
\end{equation}
The $\eta$-functions of the angular velocity and parameters of turbulence will be specified shortly.

The EMF (\ref{10}) describes an anisotropic diamagnetic pumping by rotating turbulence. The anisotropy in particular means that the direction of transport depends on the direction of the mean magnetic field $\vec B$ \citep{Oea02}. For the often assumed case of axial symmetry, the toroidal field is transported with the velocity ${\vec U}^\mathrm{t}$. This can be seen from Eq.\,(\ref{10}) if we note that the dot product of the toroidal field with vector $\vec e$ is zero. The pumping velocity for the (axisymmetric) poloidal field can be found by noticing that only the component $\tilde{\vec U}_\perp = \tilde{\vec U} - {\vec e}({\vec e}\cdot\tilde{\vec U})$ of the velocity $\tilde{\vec U}$ normal to the rotation axis contributes to the second term on the right-hand side of Eq.\,(\ref{10}). Then, for the axisymmetric case considered, this second term can be rewritten as
\begin{equation}
    \tilde{\vec U}\times{\vec e}\left({\vec e}\cdot{\vec B}\right) = \tilde{\vec U}_\perp\times{\vec B}^\mathrm{p} ,
    \label{12}
\end{equation}
where ${\vec B}^\mathrm{p}$ is the poloidal field. This field is, therefore,  transported with the velocity
\begin{equation}
    {\vec U}^\mathrm{p} = {\vec U}^\mathrm{t} - \tilde{\vec U}_\perp .
    \label{13}
\end{equation}
For the practically important case of the $\eta$-functions of Eq.\,(\ref{11}) varying only with radius $r$, the transport velocities read
\begin{eqnarray}
    {\vec U}^\mathrm{t} &=& -{\vec e}_r\frac{\partial\eta_{_1}}{\partial r},
    \nonumber \\
    {\vec U}^\mathrm{p} &=& -{\vec e}_r\left(\frac{\partial\eta_{_1}}{\partial r} -
    \sin^2\theta \frac{\partial\eta_{_2}}{\partial r}\right)
    + {\vec e}_\theta\sin\theta\cos\theta\frac{\partial\eta_{_2}}{\partial r} .
    \label{14}
\end{eqnarray}
In this equation, ${\vec e}_r$ and ${\vec e}_\theta$ are the radial and meridional unit vectors respectively. The toroidal field is pumped radially inward and the poloidal field - inward in radius and towards the equator according to the Eq.\,(\ref{14}).

\begin{figure}
\centering
   \includegraphics[width=10 truecm, angle=0]{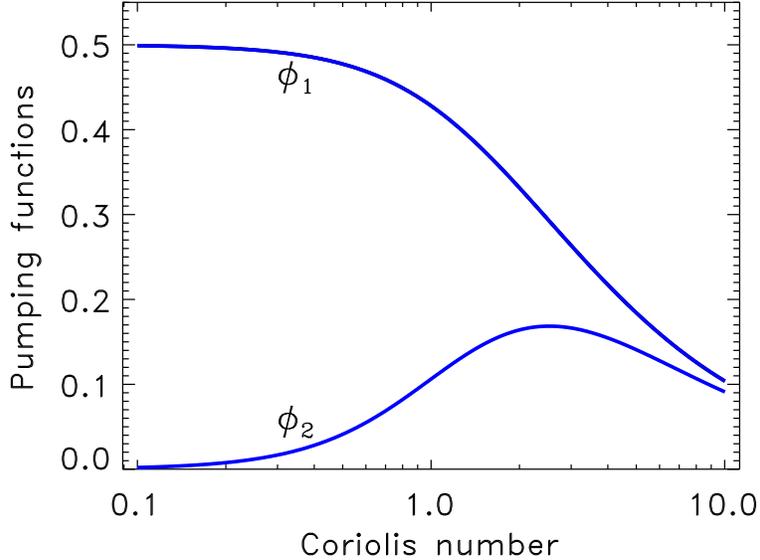}
   \caption{$\phi$-functions of Eq.\,(\ref{17}) defining the dependence of diamagnetic pumping on rotation rate.
            }
   \label{f2}
\end{figure}

Quasilinear expressions for the $\eta$-functions are rather complicated. They include spectral integrals of complicated functions \citep[cf.][]{KR92}, not appropriate for use in dynamo models. Sufficient simplification provides the mixing-length approximation (also known as the $\tau$-approximation). In this approximation, the local spectrum (\ref{9}) includes the only scale $\ell = u\tau$ ($u = \langle (u^0)^2\rangle^{1/2}$ is the rms velocity and $\tau$ is the convective turnover time):
\begin{equation}
    E(k,\omega,{\vec r}) = 2 u^2 \delta(k - \ell^{-1})\delta(\omega).
    \label{15}
\end{equation}
Viscosity and magnetic diffusivity are replaced by the order of magnitude estimations of their effective values
\begin{equation}
    \nu = \eta = \ell^2/\tau .
    \label{16}
\end{equation}
The $\eta$-functions of Eqs (\ref{11}) and (\ref{14}) then read
\begin{equation}
    \eta_{_1} = \eta_{_\mathrm{T}}\phi_{_1}(\Omega^*),\ \
    \eta_{_2} = \eta_{_\mathrm{T}}\phi_{_2}(\Omega^*) ,
    \label{17}
\end{equation}
where $\eta_{_\mathrm{T}} = \ell u/3$ is the background eddy diffusivity and the dependence on rotation rate is involved via the new $\phi$-functions of the Coriolis number
\begin{equation}
    \Omega^* = 2\tau\Omega .
    \label{18}
\end{equation}
This number measures the intensity of interaction between convection and rotation. The Coriolis number in the Sun increases with depth and exceeds 10 near the bottom of the convection zone where the diamagnetic pumping is most significant. The $\phi$-functions of Eq.\,(\ref{17}) read
\begin{eqnarray}
    \phi_{_1}(\Omega^*) &=& \frac{3}{4{\Omega^*}^2}
    \left( -1 + \frac{1 + {\Omega^*}^2}{\Omega^*} \tan^{-1}(\Omega^*)\right) ,
    \nonumber \\
    \phi_{_2}(\Omega^*) &=& \frac{3}{4{\Omega^*}^2}
    \left( -3 + \frac{3 + {\Omega^*}^2}{\Omega^*} \tan^{-1}(\Omega^*)\right) .
    \label{19}
\end{eqnarray}
The functions are shown in Fig.\,\ref{f2}.

In spite of the ${\Omega^*}^2$ in denominators of the functions (\ref{19}), they approach finite values for the small Coriolis number: $\phi_{_1} \simeq 1/2 - {\Omega^*}^2/10$ and $\phi_{_2} \simeq {\Omega^*}^2/5$ for $\Omega^* \ll 1$. In the limit of vanishing rotation, therefore, the pumping becomes isotropic with equal transport velocities ${\vec U} = -{\vec\nabla}\eta_{_\mathrm{T}}/2$ for poloidal and toroidal fields to reproduce the known result for non-rotating fluids \citep{KR80}.

In the opposite limit of very fast rotation, the $\phi$-functions of Eq.\,(\ref{19}) decrease in inverse proportion to the Coriolis number: $\phi_{_1} \simeq \phi_{_2} \simeq  3\pi/(8\Omega^*)$ for $\Omega^* \gg 1$. The pumping velocities (\ref{14}) for toroidal and poloidal fields differ in this rapid rotation case. Close to the equator, in particular, the transport velocity for the poloidal field vanishes while the toroidal field is transported with the velocity ${\vec U}^\mathrm{t} = -{\vec\nabla}\eta_{_2}$. This finding reproduces the long known result of \citet{Z57} for two-dimensional turbulence. The point here is that turbulence approaches two-dimensionality at the rapid rotation. The flow is almost uniform along the rotation axis. \citet{Z57} found that the field normal to the direction along which the 2D turbulent flow does not vary is transported with the effective velocity ${\vec U} = -{\vec\nabla}\eta_{_2}$, where $\eta_{_2}$ is the eddy diffusivity for this field. The field parallel to this direction, however, is not subject to the pumping. This is why the pumping velocity of the poloidal field disappears near the equator at the rapid rotation. It remains to be noted that the turbulent magnetic diffusion is also anisotropic in rotating fluids and the $\eta_{_2}$ diffusivity applies to the direction normal to the rotation axis in the rapid rotation limit \citep[][the diffusivity along the rotation axis is twice as large]{Kea94}.
\section{Conclusion}
Super-kilogauss toroidal fields with magnetic flux in excess of $10^{24}$~Mx in the near-bottom region of the convection zone are required in order to reverse and then replenish the global poloidal field by the Babcock-Leighton mechanism in the course of a solar cycle \citep{CH15,MT15}. Such strong fields, however, cannot be produced by the differential rotation in a distributed dynamo scenario (Fig.\,\ref{f1}). Some mechanism for the field concentration at the base of convection zone is necessary. The concentration can be provided by the diamagnetic expulsion of large-scale fields from the regions of relatively large turbulence intensity \citep{Z57,KR80}. The diamagnetic pumping is most efficient near the base of the convection zone where convective turbulence is strongly inhomogeneous. Allowance for the downward pumping helps to realise an interface dynamo in models with distributed design \citep{KO12}.

In this paper, diamagnetic pumping by rotating turbulence was derived. The transport velocities of Eq.\,(\ref{14}) reproduce previously known results for cases of very slow and very rapid rotation and extend them to arbitrary rotation rates. The transport velocities for (axisymmetric) toroidal and poloidal fields differ. The toroidal field is pumped downward while the transport of the poloidal field has a horizontal part towards the equator. The horizontal transport can influence the latitudinal migration of the field in dynamo models \citep{GG08} though the field advection by the meridional flow is probably more significant.

A more important effect of rotation seems to be the increase of the proportionality coefficient between the pumping velocity of the toroidal field and the diffusivity gradient. The increase from 1/2 for slow rotation to 1 for rapid rotation may seem to be slight but can, nevertheless, be important. Let us imagine a slab of turbulent fluid bounded by, say, horizontal superconducting planes. Let the large-scale horizontal field $B$ in the slab be subject to turbulent diffusion with diffusivity $\eta_\mathrm{t}$ and pumping with the effective velocity ${\vec U} = -q{\vec\nabla}\eta_\mathrm{t}$. If both the field and the eddy diffusivity are horizontally uniform and vary in vertical direction only, the steady solution of the mean-field induction equation demands $B\eta_\mathrm{t}^q = const$. The steady field-profile is very sensitive to the power index $q$ if the diffusivity varies across the slab as strongly as it does near the base of the solar convection zone.

Other types of turbulent transport can operate in stellar convection zones. The topological pumping of \citet{DY74} due to the asymmetry between upward and downward flows is a widely known example. However, the strong inhomogeneity of convective intensity near the base of the convection zone probably makes the diamagnetic pumping dominant in this region. It is also not clear whether the topological asymmetry can be relevant to the almost 2D convection at high Coriolis numbers.
\section*{Acknowledgement}
This work was supported by the Russian Foundation for Basic Research (project No. 16-02-00090).
\section*{References}

\end{document}